\documentclass[preprint2]{aastex6}

\usepackage{lineno}
\usepackage{graphicx}

\begin{document}


\title{Modeling the Variable Polarization of $\epsilon$~Aurigae In and Out of Eclipse}



\author{Richard Ignace, Gary Henson, }
\affil{Department of Physics \& Astronomy,
East Tennessee State University,
Johnson City, TN 37614, USA}

\author{Hilding Neilson}
\affil{David A.~Dunlap Department of Astronomy \& Astrophysics, University of Toronto, 50 St.~George Street, Toronto, ON, Canada, M5S~3H4}

\and

\author{William Asbury}
\affil{Department of Physics \& Astronomy,
East Tennessee State University,
Johnson City, TN 37614, USA}

\begin{abstract}

The eclipsing binary $\epsilon$~Aur is unique in being
a very long-period binary involving an evolved, variable F 
star and a suspected B main-sequence star enshrouded in an opaque
circumstellar disk.  The geometrical arrangement is that
the disk is viewed almost perfectly edge-on, with alignment
leading to a partial eclipse of the F~star.  Despite a global
observing campaign for the 2009-11 eclipse, there remain outstanding
questions about the nature of the binary, its components, the
disk, and the evolutionary state of the system.  We analyze
optical-band polarimetry in conjunction with broad-band
color variations to interpret brightness variations across the
surface of the F~star.  We model this both during and after
the 1982-84 eclipse for which an extensive and dense data set exists.  
We develop a model in terms of surface
temperature variations characterized by a small global
variation overlaid with a temperature variation described
with low-order spherical harmonics. While not providing a detailed
fit to the dataset, our modeling captures the overall
characterization of the color and polarimetric variability.
In particular, we are able to recover the gross behavior of
the polarimetric excursion in the Q-U plane as observed during
eclipse of the F~star when compared to post-eclipse behavior.

\end{abstract}



\section{Introduction} \label{sec:intro}

The bright eclipsing binary system $\epsilon$~Aur (HD 31964) has
the longest known orbital period (27.1 years) and longest duration
primary eclipse ($\sim 2$ years) of all eclipsing binary stars.  It
is comprised of a large F0 star and a huge dark disk,
most likely enshrouding an early B-type star, as the companion to
the F~star.  Efforts to determine the nature of the nearly invisible
disk-like companion have dominated observations of the system.
However, information about the disk is obtained only by observations
of its effects on light from the F~star during the infrequent
primary eclipses.  The most recent 2009-2011 eclipse yielded a
wealth of new results \citep[and references therein]{2015ASSL..408..107S}.
In addition to increased photometric, spectroscopic, and
spectropolarimetric monitoring, optical interferometric observations
have led to synthetic images of the disk eclipsing the large F~star
on the sky \citep{2015ApJS..220...14K}.  

There still remains a substantial uncertainty in the distance to
the system though most recent works argue for a distance less than
1~kpc \citep{2018MNRAS.479.2161G}. These latest observational constraints
confirm $\epsilon$~Aur to be an interacting binary system. However, application
of the latest stellar evolution codes for binary stars generates
models that allow for the F~star to range from a post-RGB type to
a pre- or post-AGB type star \citep{2018MNRAS.479.2161G}. The precise
nature of the binary components of this system is still under debate.


As complex as the eclipsing disk surrounding a B-type star appears
to be, interpretation of the eclipse observations remains a challenge
as the F~star is known to undergo semi-regular pulsations
\citep{1985MNRAS.216..571A, 2008JASS...25....1K, 2018AstL...44..457I}.
Spectro-interferometric observations of H$\alpha$ have also shown
it to have an extensive P-Cygni like wind region
\citep{2012A&A...544A..91M}.  The pre- and post-eclipse interferometric
data of \cite{2015ApJS..220...14K} indicate the F~star has a
more-or-less constant diameter.  However, they note some non-zero
closure phases for their baseline models, coupled with variations
in radius and limb darkening during eclipse, suggesting the possible
presence of convective cells or thermal features (e.g., ``spots'')
on the surface.  Such thermal variations may be expected to arise
from non-radial pulsations (NRPs) and to produce a small linear
polarization signature from a simple free electron scattering model
\citep{1980AcA....30..193S}. Alternatively, the thermal variations
could be represented by considering the stellar surface as radiating
anisotropically with Thomson or Rayleigh single scattering in a
spherical circumstellar envelope producing a polarization signature
from components of the spherical harmonics of such anisotropy
\citep{1999A&A...347..919A}.

Broadband polarization observations were made during the 1982-1984
eclipse \citep{1986ApJ...300L..11K} and the 2009-2011 eclipse
\citep[]{2012JAVSO..40..787C,2012AIPC.1429..140H}.  A significant
polarization signature is produced by the extreme asymmetry of the
eclipse geometry. However, \cite{1989PhDT........11H} proposed that
NRPs of the F star create variable polarization with an amplitude
similar to the effect of the eclipse and with a typical timescale
on the order of 100 days. His work focused on the variability outside
eclipse beyond the 1982-1984 event. Neither the \cite{2012JAVSO..40..787C}
nor the \cite{2012AIPC.1429..140H} observations were of the quality
or density of those made by \citep{1986ApJ...300L..11K} and
\cite{1989PhDT........11H}.  \cite{2012JAVSO..40..787C} did report
variable polarization outside the eclipse similar in scale to that
of \cite{1989PhDT........11H}, but his observations spanned only
$\sim$150 days. Although \cite{2012AIPC.1429..140H} observed several
months pre-eclipse, the limited precision of the observations did
not allow the detection of any significant variations outside of
eclipse. As each of the above authors chose to analyze the eclipse
behavior using $q$ and $u$ Stokes parameters with different coordinate
reference frames (i.e., rotating the normalized parameters by
different angles), a direct comparison of the eclipse polarization
cannot be easily made \citep[for an overview of Stokes parameters
and polarimetry see e.g.,][]{2010stpo.book.....C}.  However, a
simple qualitative comparison near the stages of egress (the only
phase common to all observers) does show the polarization in both
parameters changing significantly as the geometry of the system
experiences its most dramatic change.

Broadband photometric variability is also present outside the
eclipse. Persistent $\sim$0.1 magnitude variations exist on a
timescale of $\sim$2 months, but both the amplitude and period vary
unpredictably.  This behavior, along with the elongated disk geometry,
complicates the establishment of the contact points for the eclipse
\citep{2012JAVSO..40..668K}.  In addition, \cite{2012JAVSO..40..668K}
points out that ingress and egress have different lengths and slope
variations. Thus, although the amplitude of the 
photometric variability outside eclipse is much smaller than the eclipse depth, its
irregularity makes it harder to clearly define the features
of the eclipse itself. For the polarization, it was noted earlier
that the amplitude of the polarized light variations outside eclipse 
is significant when compared to the overall eclipse effect. Since
they also vary unpredictably in amplitude and period, it is extremely
difficult to separate eclipse effects from the behavior of the
variable F star.

We construct here a simple model for a polarized atmosphere for the
F0 star.  A phenomenological model of surface brightness
variations is introduced in an effort to characterize the aggregate
properties observed in the variable polarization both in and out
of eclipse for the star.  In \S~\ref{sec:obs}, a review of the
dataset used for comparison with the modeling is described.  The
description for our model of variable polarization is given in
\S~\ref{sec:atm}.  Results from the modeling are presented in
\S~\ref{sec:results}, with a summary of this study and concluding
remarks given in \S~\ref{sec:summary}.

\section{Observations}  \label{sec:obs}

The $V$ band linear polarization measurements of $\epsilon$~Aur
modeled here were obtained during and after the 1982-84 eclipse by
\cite{1989PhDT........11H}.  These comprise an extensive
and precise set of measurements consisting of 969 nightly data
points spanning from 1982 August to 1988 May (representing
approximately 20 cycles of a generic pulsation period on the order
of 100 days).  The measurements were obtained using the
photoelastic-modulator polarimeter attached to the 61~cm telescope
at the University of Oregon's {\em Pine Mountain Observatory} described
in detail by \cite{1981PASP...93..521K}.  The dataset consists
of instrumental $q$ and $u$ Stokes parameters given in percentage
polarization with typical measurement errors of only 0.010\% in the
$q$ parameter and 0.015\% in the $u$ parameter, with an instrumental
polarization for the telescope/polarimeter system of less than
0.005\%.  Thus, these errors are an order of magnitude smaller than
the typical 0.10\% fluctuation from the mean $q$ and $u$ polarization
values outside of eclipse.

\begin{figure*}
\plotone{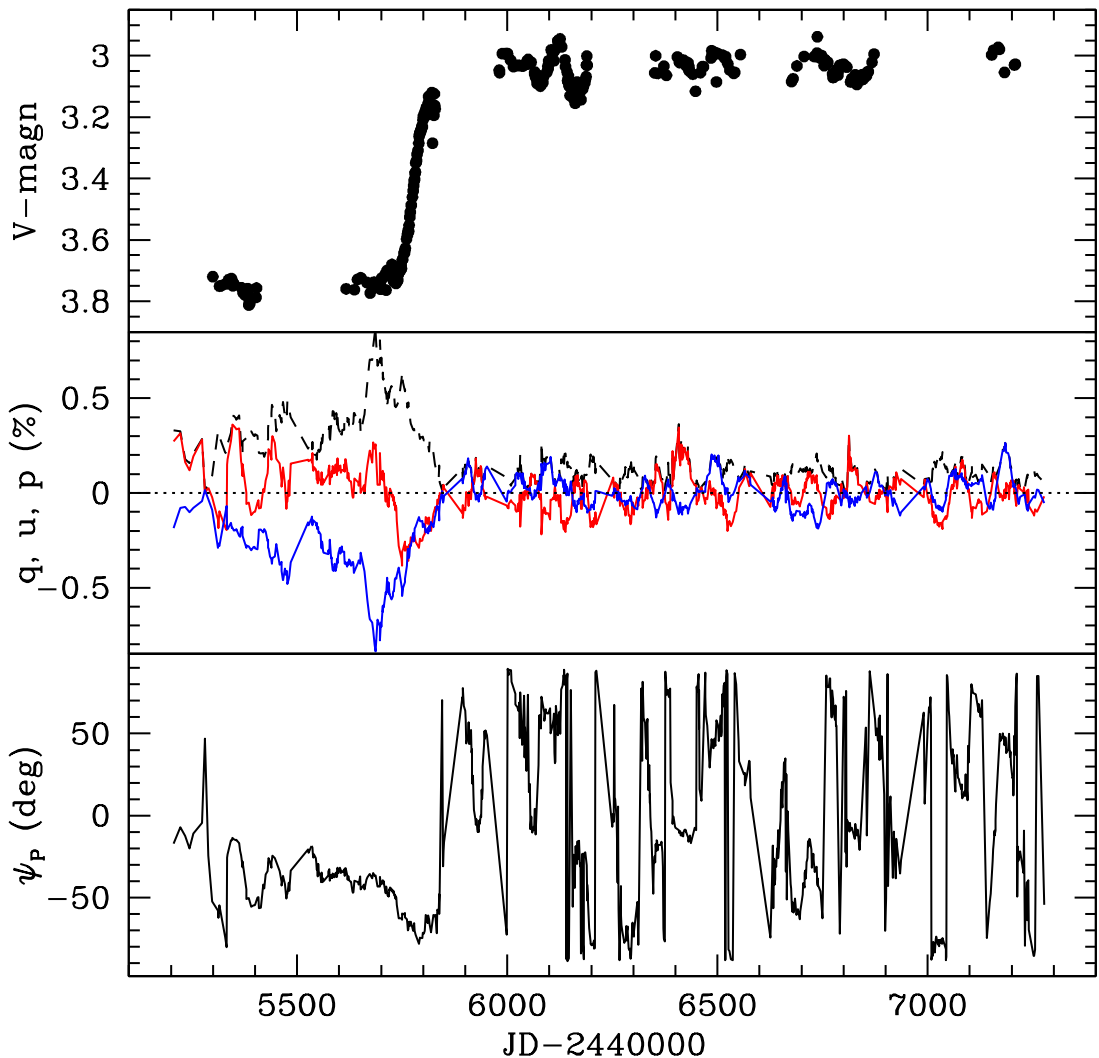}
\caption{Shown are data from the 1980's eclipse, with V magnitude
at top, polarization amplitude at middle, and polarization position
angle at bottom.  The light curves are plotted against modified
Julian date (JD) as indicated.  For the middle panel, the dashed
line is the total polarization $p$, red is $q$, and blue is
$u$.  All polarizations are indicated as percent.
\label{fig1}}
\end{figure*}

In order to facilitate comparison to the model, the instrumental
Stokes parameters were adjusted in two ways.  
First, a sizeable but
constant interstellar component of $\sim 2\%$ is present for the
measured polarization
\citep{Geise2015}.  Since the model would produce zero polarization
for a symmetric geometry, the mean value established outside the
eclipse was subtracted from each parameter.  The remaining variability
would thus be intrinsic to the pulsations of the F~star.  Second,
as the binary system is seen edge-on to the orbital plane, the
eclipse aspects of the model comparison can best be facilitated if
the measured Stokes parameters are oriented relative to the orbital
plane rather than presented as the standardized equatorial parameters
which would align the $q$ parameter with north and south on the
sky.  We benefit from the interferometry results for the 2009-2011
eclipse \citep{2015ApJS..220...14K} which establish a position angle
of $297^\circ \pm 7^\circ$ on the sky for the disk-orbit plane.
Thus, the original instrumental parameters were rotated with a
Mueller matrix (which preserves the total polarization) by an angle
of $-27^\circ$ to align the $q$ parameter perpendicular to the orbit
plane.  Figure~\ref{fig1} shows V band photometry from the
AAVSO\footnote{We acknowledge with thanks the variable star
observations from the AAVSO International Database contributed by
the Hopkins Phoenix Observatory and used in this research.}, the
$q$ (red) and $u$ (blue) parameters, the percent polarization (black;
$p=\sqrt{q^2+u^2}$), and the polarization position angle $\psi_P$
versus time for the rotated, normalized polarization measurements
that we compare to the model results.

\begin{figure}
\plotone{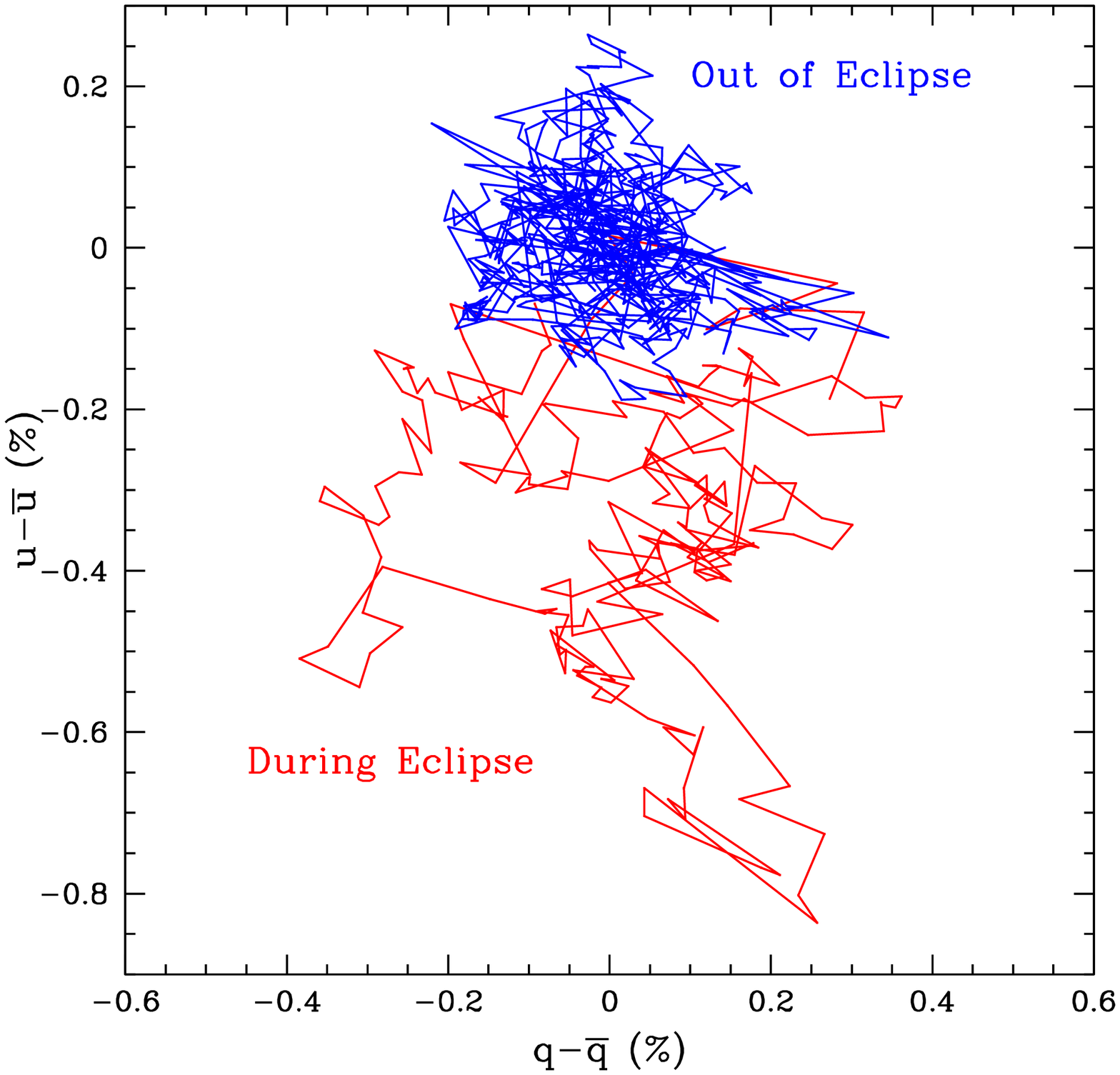}
\caption{Here the $q$ and $u$ data from Fig.~\ref{fig1} are plotted
in the $q-u$ plane.  As labeled, the red portion is during eclipse,
showing an excursion from the blue portion which is post-eclipse.  The
polarization is plotted relative to centroid values $\bar{q}$ and
$\bar{u}$ associated with the post-eclipse data, assuming those
average values represent the interstellar polarization and
any stable intrinsic polarization to the binary system.
\label{fig2}}
\end{figure}

\section{Model for Stellar Polarimetric Variability} \label{sec:atm}

\subsection{Description of the Model}

The intrinsic and variable polarization of the F~star in $\epsilon$~Aur
offers a number of challenges to modeling efforts.  Owing to its
low gravity, one needs to calculate polarized radiative transfer
for a model atmosphere in spherical coordinates to account for
extended atmosphere effects \citep{1971MNRAS.154....9C,
1986ApJ...307..261D, 2015A&A...575A..89K, 2016A&A...586A..87K}.
Plus, the atmosphere must be time-dependent and 3-dimensional to
capture fully and accurately the variable and asymmetric behavior
required to model the observations.  Prior to investing effort in
such a challenging undertaking, it is prudent first to employ
phenomenological models to anticipate trends that may be expected
from detailed atmosphere calculations \citep[e.g., ]{1999A&A...347..919A,
2009A&A...496..503I}.  Such an approach affords a more rapid
exploration of a large parameter space to obtain a basic interpretation
of observations.

We have constructed a simple model for a polarized atmosphere.  At
early F~spectral type, the star is hot enough that dust and molecular
opacities are negligible.  The primary polarigenic opacity will be
electron and Rayleigh scattering \citep{1972ApJS...24..255O}, both
of which are dipole scattering mechanisms.  In summary, our model
adopts the following assumptions:

\begin{itemize}

\item We choose a parametric center-to-limb profile for the variation
of the polarization of light to represent a static and spherically
symmetric atmosphere.  The selection involves two considerations:
a shape function with impact parameter for a distant observer, and
an amplitude value at the stellar limb.

\item The star is assumed spherical in shape at all times at fixed
radius; however, we allow for brightness variations across the
surface of the star.  We choose to parametrize the variations in
terms of temperature with position across the atmosphere.  The
manner for achieving this involves two considerations: a pattern
function across the stellar surface, and an amplitude for the
temperature variation.  A variation of temperature that maintains
spherical symmetry (e.g., radial pulsations) has no effect on the
net polarization for an unresolved source, although it does alter
the color (i.e., $B-V$) of the star.  Consequently, some ``patchy''
description for brightness variations is needed to break spherical
symmetry, as might arise from a distribution of convective cells
of varying brightness's. Based on the interferometric studies along
with the semi-regular nature of the F~star component, such a patchy
description seems reasonable.  For a surface pattern function, we
will consider low-order modes of spherical harmonics to represent
the temperature variations that determine the brightness variations.

\item The hydrogen absorption lines strengthen through the F and
into the A spectral type.  Consequently, with temperature variations
are associated variations in the opacity with depth in the atmosphere,
including ionization of hydrogen, which affects the electron number
density in the atmosphere.  To include this influence, the Saha
equation is used as a simple means of modulating the polarigenic
opacity as a function of the position-dependent temperature through
changes in the ionization.

\item All stars rotate.  Adopting a fixed pattern of temperature
variations, stellar rotation leads to a phase drift for how the
temperature variations appear in the light curve, including the
polarization properties.  A $v\sin i = 38$ km s$^{-1}$ has
been measured for the F~star \citep{2012Ap.....55..528P}.  Ambiguity
in the distance leads to an uncertain stellar radius and rotation
period.  However, a rotation period of order $10^2$~d is reasonably
expected.  Here we treat the period as a free parameter of the
model, assuming solid-body rotation of the atmosphere.


\end{itemize}

\begin{figure}[ht!]
\plotone{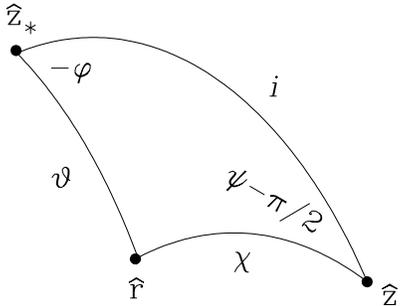}
\caption{Relationship between the observer and stellar coordinates, as described in the text.
\label{fig3}}
\end{figure}

\subsection{Specification of the Model}

Figure~\ref{fig3} defines the geometry adopted for our model in terms of a spherical triangle.  The center of the sphere
is the center of the star.  The unit vector $\hat{z}_\ast$ specifies
the spin axis of the star.  The unit vector $\hat{z}$ specifies the
direction of the observer from the star.  The spin axis is viewed
at an inclination angle, $i$, given by $\cos i = \hat{z} \cdot
\hat{z}_\ast$.

The unit vector $\hat{r}$ signifies a radial from the star to a
point in the stellar atmosphere. The spherical angular coordinates for the
star system are $(\vartheta,\varphi)$, as indicated, where $\vartheta$ is
a co-latitude.  The spherical angular coordinates for the observer system
are $(\chi,\psi)$.

Model fluxes can be computed for known intensities as follows.  We use Stokes
parameters $I,Q,U,V$ for our representation of the partially
polarized radiation.  However, the $V$-intensity will not be
considered further, as that parameter encapsulates information about
circular polarization, whereas our model involves linear polarization
only.

Model fluxes, $f_I, f_Q, f_U$ are computed via integral relations,
with

\begin{eqnarray}
f_I & = & \frac{R_\ast^2}{d^2}\,\int I_\nu(\chi,\psi)\,\sin \chi\,d\chi\,d\psi, \\
f_Q & = & \frac{R_\ast^2}{d^2}\,\int Q_\nu(\chi,\psi)\,\sin \chi\,d\chi\,d\psi, \\
f_U & = & \frac{R_\ast^2}{d^2}\,\int U_\nu(\chi,\psi)\,\sin \chi\,d\chi\,d\psi.
\end{eqnarray}

\noindent As noted in the previous section, instead of obtaining
the intensities from a detailed atmosphere calculation, a parametric
approach is adopted.  For a spherically symmetric star, the intensity
with angular radius $\chi$ is prescribed as a linear limb-darkening
law:

\begin{equation}
I_\nu = (a_\nu+b_\nu\cos\chi)\,B_\nu(T).
\end{equation}

\noindent where $B_\nu$ is the Planck function, $T$ is the effective
temperature at location $(\chi,\psi)$, and the parenthetical accounts
for the limb darkening.  We adopt coefficients of $a_\nu=0.4$
and $b_\nu=0.6$ from the plane-parallel grey atmosphere solution with the Eddington 2-stream approximation as being
adequate for our phenomenological approach.  There are numerous
papers that discuss detailed calculations and parametric formulae
for wavelength-dependent limb darkening \citep[e.g., ]{2013A&A...554A..98N, 2013A&A...556A..86N}, and while we recognize that sphericity effects are likely important for limb darkening of the F~star, the goal here is
to include only basic ingredients of the atmospheric physics to
achieve qualitative trends, and linear limb darkening is adequate
to this purpose.

Next we approximate the emergent monochromatic polarized
intensities with:

\begin{equation}
Q_\nu = (a+b\cos\chi)\,B_\nu(T)\,p_{\cal L}(T)\,(1-\cos^2\chi)\,\cos 2\psi,
	\label{eq:Qnu}
\end{equation}

\noindent and

\begin{equation}
U_\nu = (a+b\cos\chi)\,B_\nu(T)\,p_{\cal L}(T)\,(1-\cos^2\chi)\,\sin 2\psi.
	\label{eq:Unu}
\end{equation}

\noindent Note that we drop the frequency subscript from the limb
darkening coefficients as these are assumed constant.  In equations~(\ref{eq:Qnu}) and (\ref{eq:Unu}), the second
parenthetical of $(1-\cos^2\chi)$ is the shape function we choose to
represent how the polarized intensity varies with angular radius
from the center of the star as seen by the observer.  It is a
parabolic function, with zero polarization from the direction
of star center, and maximum polarization at the star's
limb\footnote{Although the shape function for the polarization
maximizes the relative polarization at the limb, it does not follow
that the limb is the greatest contributor to the net polarized flux,
owing to the effects of limb darkening.}.  \cite{1970Ap&SS...8..227H} has
calculated the center-to-limb emergent intensity polarization profiles
for grey atmospheres.  That paper also presents approximation
formulae for the results, and while having a profile shape that is
more complicated than our simple parabola, the leading term for the
approximation is a parabola, which we consider adequate for our
purposes.

The amount of limb polarization for this function is specified by
the amplitude $p_{\cal L}(T)$ at the star limb, which we take as a
function of temperature, with:

\begin{equation}
p_{\cal L} = p_0\,\left(\frac{T}{7400~{\rm K}}\right)^{10}.
\end{equation}

\noindent A scaling for the limb polarization with temperature was
guided from use of the Saha equation. The latter is adopted
for how the relative ionization of H in the atmosphere
depends on temperature.  Our use of the Saha equation is basic, but
it does provide a convenient estimate for how the scattering opacity
responds to changes in local temperature.  Assuming an atmosphere
of pure H, let the number density of hydrogen be $n_H = n_+ +
n_0$, where $n_0$ is the number density of neutral hydrogen, and
$n_+$ is for  ionized hydrogen.  We introduce $\epsilon_+ =
n_+/n_H$ and $\epsilon_0 = n_0/n_H$ as fractions, such that $\epsilon_+
+ \epsilon_0 =1$.  Finally, the number density of electrons is given
by $n_{\rm e} = n_+$.  The Saha equation yields a well-known
expression of the form,

\begin{equation}
\epsilon_+^2 + \Gamma\,\epsilon_+ - \Gamma = 0,
\end{equation}

\noindent where

\begin{equation}
\Gamma(T,P_H) = \left(\frac{2\pi\,m_{\rm e}}{h^2}\right)^{3/2}\,
	\frac{(kT)^{5/2}}{P_H}\, e^{-E_0/kT},
\end{equation}

\noindent in which $E_0$ is the ionization potential of hydrogen,
$k$ is the Boltzmann constant, $h$ is the Planck constant, $m_{\rm e}$
is the electron mass, and $P_H$ is the hydrogen gas pressure.

At around 7,000~K, and $\log g \approx 1.5$, we estimate $\kappa
\approx 0.4$ cm$^2$ g$^{-1}$ and derive a gas pressure at $\tau =
2/3$ of $P_H \approx 70$.  We estimate this based on opacities for
an atmosphere model by \cite{1972ApJS...24..255O} combined with the
pressure distribution for a gray atmosphere.  Curves for $\epsilon_+$
can be computed as functions of $T$.  In the range of 6,800--8,300~K,
a power-law fit to the curve can be obtained, which is given
approximately by

\begin{equation}
\epsilon_+ \propto T^{10}.
\end{equation}

\noindent The temperature range of applicability for
the preceding relation is roughly the maximum variation
expected from our models.  The steep power-law represents the
fact that the solution for ionization of H is sensitive to
the exponential in the Saha equation in this temperature range.

For the variation of temperature across the star, we employ 
the following expression:

\begin{equation}
T^4 = T^4_0\,\left[1+\gamma_0\,\cos(\Omega_0 t)\right] 
	+ \Delta T^4\,G(t,\vartheta,\varphi).
	\label{eq:T4}
\end{equation}

\noindent Here the function $G$ represents the pattern of temperature
variations across the stellar surface.  It is defined in such a way
that integration of $G$ across the surface is zero, hence $T_0$ is
the time-averaged temperature.  Then $\Delta T$ is treated as a
free parameter for the amplitude of temperature variations.  

The average surface temperature at any given time is allowed to
vary sinusoidally.  The parameter $\gamma_0$ is an amplitude
for this global variation.  The angular frequency $\Omega_0 =
2\pi/P_0$, for a period of temperature variation $P_0$.
As will be seen, some variation in average temperature is needed
to match color variations observed in $\epsilon$~Aur.  A reminder that
$\gamma_0$ and $\Omega_0$ make no contribution to polarization of
the star, variable or otherwise, since these relate to the average
spherical star; polarimetric variability arising only from the
second term of equation~(\ref{eq:T4}).

For $G$ to describe the brightness variations, we use a prescription based on spherical harmonic functions with

\begin{equation}
G = \sum w_{l,{\rm m}}\,\tilde{Y}_{l}^{\rm m}(t,\vartheta,\varphi) 
	\,\sin(\Omega_{l}t) / \sum w_{l,{\rm m}}.
\end{equation}

\noindent Here, $\tilde{Y}_{l}^{\rm m}$ are the standard spherical
harmonic functions with the leading constants omitted.  For example,
$Y_1^0 = \sqrt{3/4\pi}\,\cos\vartheta$, for which $\tilde{Y}_1^0 =
\cos\vartheta$.  The coefficients $w_{l,{\rm m}}$ are weights for the summation
of terms, and are treated as free parameters.  Time dependence enters the
function through rotation of the pattern about the star's spin axis, modulating
the temperature variations in an oscillatory manner.  We also allow for the possibility of ``mode switching'' (to be discussed later).  The 
modulation is made explicit via the sinusoidal time dependence
in terms of $\Omega_{l} = 2\pi/P_{l}$, where $P_{l}$ represents a
characteristic timescale for the pattern of temperature
variations associated with order $l$.

The function $G$ for temperature variations across the star is expressed
in terms of the stellar coordinates, $(\vartheta, \varphi)$.
However, the $\epsilon$~Aur system is eclipsing and thus seen near edge-on
to the orbital plane.  For a general viewing inclination, the relations
between stellar and observer coordinates can be determined from
spherical trigonometry, to yield:

\begin{eqnarray}
\cos \chi & = & \cos i\,\cos \vartheta + \sin i \, \sin \vartheta\,
	\cos \varphi,  \\
\tan \psi & = & \frac{\sin i\,\cos \vartheta - \cos i \, \sin \vartheta
	cos \varphi}{\sin \vartheta\,\sin \varphi}, \\
\cos \vartheta & = & \cos i\,\cos \chi + \sin i \, \sin \chi \,\sin \psi, \\
\tan \varphi & = & \frac{\sin \chi \, \cos \psi}{\cos \chi \,\sin i -
	\sin \chi \, \sin \psi}.
\end{eqnarray}

\noindent In the special case of $i=90^\circ$, the relations reduce to:

\begin{eqnarray}
\cos \chi & = & \sin \vartheta \,\cos \varphi, \\
\tan \psi & = & \left(\sin \varphi\,\tan\vartheta\right)^{-1}, \\
\cos \vartheta & = & \sin \chi \, \sin \psi, \\
\tan \varphi & = & \tan \chi \, \cos \psi.
\end{eqnarray}

Once a structure function $G$ has been chosen (i.e., values of $l$, $m$, and the weights), along with the other
free parameters (e.g., $\gamma_0$, $\Delta T^4$, $T_0$, and $\Omega_0$), the integral
relations for the Stokes fluxes can be numerically evaluated for
each time step to produce simulated light curves.  Those fluxes
can be combined to yield normalized $q$ and $u$ parameters, as
well as total polarization $p$ and net polarization position
angle $\psi_P$, for the system as functions of time.  The relations are

\begin{eqnarray}
q(t) & = & f_Q/f_I, \\
u(t) & = & f_U/f_I, \\
p(t) & = & \sqrt{q^2+u^2},~{\rm and} \\
\tan 2\psi_P(t) & = & u/q. \\
\end{eqnarray}

\noindent In addition monochromatic magnitude variations are computed as

\begin{equation}
\Delta m(t,\lambda)= -2.5 \log \left[\frac{f_I(\lambda)}{f_0(\lambda)}\right],
\end{equation}

\noindent where

\begin{equation}
f_0(\lambda) = \int (a+b\cos\chi)\,B_\nu(T_0)\,\sin \chi\,d\chi\,d\psi. \\
\end{equation}

\noindent Here $f_0$ is the stellar flux evaluated at the time-averaged
temperature $T_0$.  Note that the coefficients $\{w_{\rm l,m}\} = 0$ except when $l$ and $m$ are specified modes for a simulation.

\begin{figure}
\hspace{-1in}\includegraphics[width=\columnwidth]{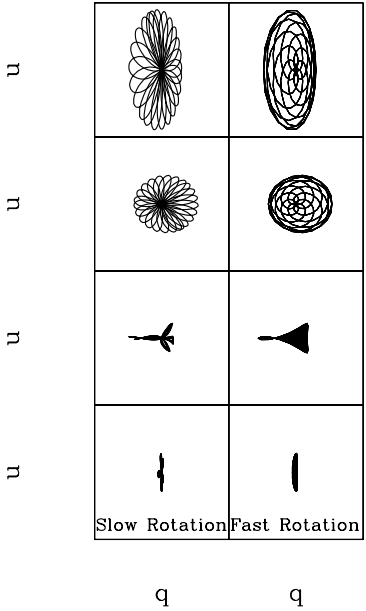}
\caption{Polarization:  top to bottom is $(l,m) = (2,1), (2,2), (3,1),
(3,2)$.  Left is slow rotation; right is fast rotation.
\label{fig4}}
\end{figure}

\begin{figure}
\hspace{-1in}\includegraphics[width=\columnwidth]{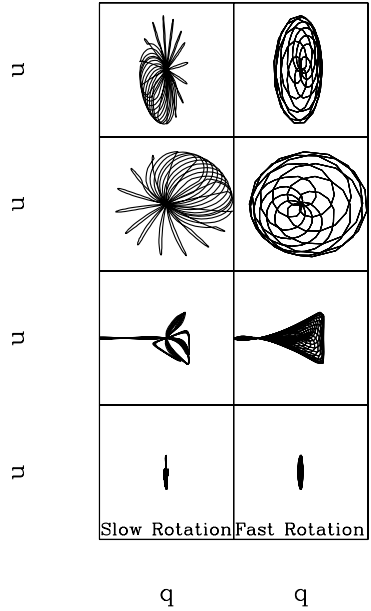}
\caption{Polarization as in the previous figure, but now with a beat frequency for a phase
shift (see text for further description).
\label{fig5}}
\end{figure}


\subsection{Illustrative Examples}

This section presents a few example calculations to illustrate the
output of the model.  Two versions are presented.  In the first a
single $(l,m)$ mode is considered for the pattern of temperature
variations in conjunction with stellar rotation.  Figure~(\ref{fig4})
shows model results in plots of $q$ versus $u$.  We adopt $P_2$ as
the ``generic'' timescale of variations in the F~star, and set $P_2
= 0.3$~years.  Other periods are referenced to this value.  The
simulations in this figure run for $12\times P_2=3.6$ years.  The
left column panels are for a relatively slow rotation with $P_\ast
= 3.3 P_2$; the right column is for a relatively fast rotation with
$P_\ast = 0.3 P_2$.  For $l=3$, we use $P_3 =  2/3 \times P_2$.
The modes from top to bottom are:  (2,1), (2,2), (3,1), and (3,2).

In each case the viewing inclination is $i=90^\circ$, and the
temperature amplitude is $\Delta T^4 = 0.25 T_0^4$.  All of the
examples adopt $\gamma_0 = 0$, hence there is no time-dependence
of $T_0$, which is fixed at $T_0=7,400$~K, the nominal temperature
of the F~star. The variation in $T$ across the star ranges from
roughly 6900~K up to 7800~K, as characterizing the brightness
variations.  The time-dependence of these temperatures is modulated
according to $\sin(\Omega_{\rm l}\, t)$, where $\Omega_{\rm l} =
2\pi/P_{\rm l}$.  At the same time, the pattern of brightness
variations rotates according to $\sin \Omega_\ast t$.  The examples
of Figure~\ref{fig4} all use the same value of $P_2$.  The $l=2$
modes tend to be oval in shape.  For $l=3$, the amplitude of the
variation in the polarization pattern tends to be smaller, overall.
This is reasonable as higher modes imply more complex patterns,
leading to greater polarimetric cancellation.

Figure~\ref{fig5} displays results for the same cases of
Figure~\ref{fig4}, in the same manner, but now with inclusion of
mode switching.  For a fixed value of $l$, the value of $m$ drifts
between $+1$ and $-1$ at the angular beat frequency of $\omega_{\rm
B,l} = \| \Omega_{\rm l}-\Omega_\ast \|$.  This drift is treated
as a variation in the phase for the azimuthal dependence of the
function $G$, arising through the parameter $\phi_{l}(t)$.  We adopt
a triangular wave function in which $\phi_{l}$ varies linearly in
time from 0 to $\pi/2$, returning to 0, as given by:

\begin{equation}
\phi_l(t) = \frac{\pi}{4}\,\left\{ 1+\frac{2}{\pi}\,
	 \sin^{-1} \left[\sin(\omega_B t)\right] \right\} .
\end{equation}

\noindent The appearance of this expression may seem odd;
however, computationally, the inverse sine function conveniently returns
values in the interval of $[-\pi/2,+\pi/2]$.  

The cases in Figures~\ref{fig4} and \ref{fig5} demonstrate that
complex $q-u$ behavior can be obtained from our model atmosphere
with brightness variations governed by
spherical harmonics.  However, a single mode, even with rotation
as well as mode switching, clearly leads to orderly patterns that
are not observed in the data.  The next section presents a model
that contains more than one mode and can account for the broad
features of the observations both during and out of eclipse.

\section{Results} \label{sec:results}

The goal of our study is to determine, in principle, if the gross 
properties of the variable polarization observed in the F~star
component can be reproduced with a model of characteristic
and time-varying brightness variations across the stellar surface.
It is not our intention to fit the data, but rather to match the
characteristic behavior displayed in the data.  To that end, we
first review the observed variations.

\begin{figure}
\plotone{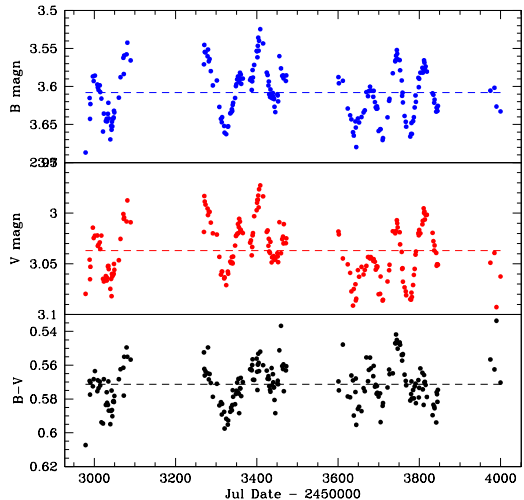}
\caption{
Model constraints for the primary of $\epsilon$~Aur involve both
characteristics of the polarization variability as well as color
variations.  Shown here are $B$ and $V$ magnitudes at top and middle,
with the $B-V$ color at bottom, used to characterize the color
variations.
\label{fig6}}
\end{figure}

\subsection{The Intrinsic Stellar Variability}

The F~star is a large and luminous star that is observed to
vary outside of eclipse.  As previously noted, it 
is a semi-regular variable with a 
characteristic period of about 100 days.  In the V-band, the variations
are of order a tenth of a magnitude relative to the mean.  Those
variations are accompanied by changes in $B-V$ color as well.
Using data taken from the AAVSO\footnote{AAVSO is the acronym for the American Association of Variable Star Observers.}, Figure~\ref{fig6} shows B and V band light curves, along with the
variations in $B-V$, out of eclipse, and over about a 3-year span of
time.  The variations in $B-V$ have a standard deviation of $\sigma(B-V)
= 0.12$.

From Figure~\ref{fig1}, the polarization out of eclipse is variable,
with a time average of $\langle p \rangle = 0.10\%$ over a span of
about 4 years.  During eclipse, the polarization is variable in a
way that consists of the effect of the eclipse of the F~star
by the disk of the secondary\footnote{Eclipse of a spherically symmetric and
time-independent atmosphere would also yield a time varying
polarization owing to the fact that there is no longer symmetry of
the projected star.  However, that varying polarization would not
exist outside of eclipse, and during eclipse, it would vary in a
smooth fashion over time.} along with the intrinsic variability of
the F~star.  In addition, the $u$ light curve remains negative
throughout the eclipse event, whereas the $q$ light curve tends to
be positive, although it can change sign.  In particular, the $q$
light curve is seen to become negative and remain negative during
egress.  The observed variations in $q$ and $u$ are reflected in
the variation of the polarization position angle in the lowermost
panel of Figure~\ref{fig1}.  There, $\psi_P$ is negative, hovering
around $-50^\circ$, throughout the eclipse, and then becomes more
randomized, with occasional rapid position angle changes, outside of
eclipse.

The dispersion in the color of the F~star, the average level
of polarization outside of eclipse, and the distinctly
different character of
the variable polarization during eclipse as opposed to outside of
eclipse are all features that we seek to reproduce with our model
involving surface brightness variations.  Note that our assumption is that
neither the edge-on disk nor the secondary star contribute to
the polarimetric variability.  If either or both are sources of
polarization, our modeling implicitly assumes those contributions
are constant over the timespan of interest, and thus subsumed
in the terms $\bar{q}$ and $\bar{u}$ as constant offsets for
the variable component of the polarization.

\subsection{Modeling Trends Outside of Eclipse}

Recalling the illustrative examples, it is clear that a function
$G$ involving only a single mode will not reproduce the observations
for polarimetric variability outside of eclipse.  A single mode
yields a pattern that is too organized in a $q-u$ plot, whereas the
observations lack such organization.  The observations even lack a clear
favored axis for the jumbled $q-u$ variations.  As a result, we attempted
to reproduce the observations using a combination of two modes.  A good
qualitative result was achieved with an equal combination of the (2,2)
and (3,1) modes, with $G = 0.5\,[G(2,2) + G(3,1)]$.  This combination
produces a pattern of $q-u$ variations that, although not a fit to the
data, are similar in character.

\begin{figure}
\plotone{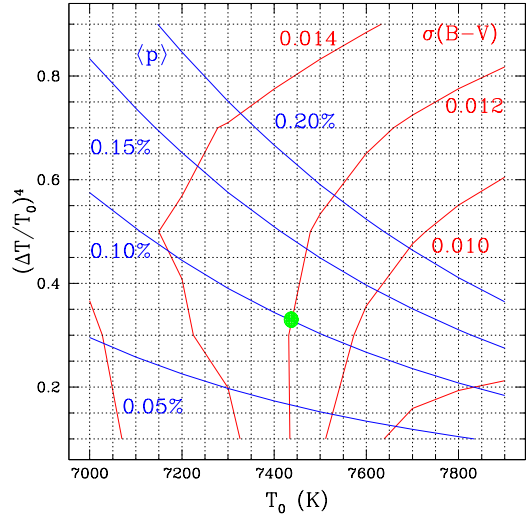}
\caption{The model space of average temperature $T_0$ and amplitude
for temperature perturbation $\Delta T$ with contours for the time
variations in color $\sigma(B-V)$ (shown as red) and for the
time-averaged polarization amplitude $\langle p \rangle$ (shown as
blue).  The green dot marks the location consistent with observations
of $\epsilon$~Aur.
\label{fig7}}
\end{figure}

Given those two modes, the next task was to simultaneously satisfy
both the observed color variations and the observed average level
of polarization.  Figure~\ref{fig7} shows a grid of models use in
determining the temperature parameters that satisfy the constraints.
The figure is a plot of $T_0$ against $\Delta T^4/T_0^4$.  The red
contours are for $\sigma (B-V)$ with levels indicated next to each
curve.  The relevant contour for $\epsilon$~Aur is the one marked
0.012.  The blue contours are for $\langle p \rangle$, again
with levels indicated.  The one relevant to $\epsilon$~Aur is 0.1\%.
The green circle indicates the intersection of the two relevant curves.

That solution has $\Delta T^4/T_0^4 =
0.33$ at $T_0 \approx 7450$~K, the latter being close to the effective temperature of $7,395\pm 70$~K found by \cite{2014AN....335..904S}.  Our result is specifically for a viewing inclination
that is edge-on to the equator of the F~star with $i=90^\circ$.  The model adopts a limb polarization
value of $p_{\cal L} = 3\%$ at a temperature of 7400~K.  A plot of
the $q-u$ variations out of eclipse, that will be shown in conjunction
with the variations during eclipse (see next section, and
Fig.~\ref{fig8}), were found to mimic the overall character of the
observations.

For the variations in $B-V$, absolute magnitude $M_B$ and $M_V$
were used as functions of temperature from \citep{2000asqu.book.....C}.
A tabulation of $M_V$ and $B-V$ as functions of temperature for
supergiant stars were used to construct intensities $I_B(T)$ and
$I_V(T)$.  At each time step in the model, the position-dependent
intensities were integrated to form net fluxes $f_B(t)$ and $f_V(t)$,
that were then combined to obtain colors $B-V$.  Then $\sigma(B-V)$
was found from the simulated color curve.  The set of contours shown
in Figure~\ref{fig7} are for $\gamma_0 = 0.08$.

Note that the solution indicated in Figure~\ref{fig7} is not unique.
For a given pattern function $G$, the contours for $\sigma(B-V)$ can
be changed by altering the choice of $\gamma_0$.  It is true that the 
effective temperature is constrained by observations at around 7,400~K;
however, the dispersion in $B-V$ can be increased by increasing $\gamma_0$.
Similarly, the contours for $\langle p \rangle$ can be shifted by
altering the choice of $p_{\cal L}$.  At a fixed value of $\Delta
T^4/T_0^4$, the value of $\langle p\rangle$ will increase as $p_{\cal
L}$ is increased.  Naturally a unique solution is preferred.  At this
stage we
may need to be satisfied with a plausible solution.  
That plausibility can be strengthened
by considering whether the chosen parameters of
the model can also explain the polarimetric behavior during the eclipse
by the secondary's disk, the topic of the next section.

\subsection{Modeling The Eclipse Event}

\cite{2015ApJS..220...14K} have conducted an extensive study of the
occulting disk properties based on optical and infrared interferometric
studies of $\epsilon$~Aur.  Their disk model involves vertical and
radial scale heights for the disk structure, along with opacities
to compute absorbing optical depths as the eclipse evolves.  Our
study for the polarimetric variability during eclipse is less
detailed and more proof-of-concept.

\begin{deluxetable}{lc}
\tablecaption{Properties of $\epsilon$ Aur\label{tab1}}
\tablehead{
\colhead{Property} & \colhead{Value} 
}
\startdata
Primary$^a$ & F0\\
Secondary$^a$ & BV(?) \\
$R_1/d^a$ & $1.11 \pm 0.05$ mas \\
$a/d^a$ & $31.3 \pm 3$ mas \\
$h_D/d^a$ & $1.04 \pm 0.14$ mas \\
$r_D/d^a$ & $7.42 \pm 0.28$ mas \\
$e^b$ & $0.23 \pm 0.01$ \\
$P_{\rm orb}^b$ & $9896 \pm 1.6$ days \\
$i^a$ & $89^\circ \pm 1^\circ$ \\
\enddata
\vspace{.2 em}
\begin{center}
{\small $^a$ from \cite{2015ApJS..220...14K}\\
\noindent $^b$ from \cite{2010AJ....139.1254S} } 
\end{center}
\end{deluxetable}

We adopt a simplified disk model for the eclipse consisting of a
rectangular shape in projection.  The disk has a long length in the
orbital plane of $16R_1$, where $R_1$ is the radius of the F~primary
star.  The full height of the rectangle can be treated as a free
parameter of the model, but is reasonably constrained from the
interferometric data during eclipse at about $1R_1$
\citep{2015ApJS..220...14K}.  All rays that intercept the disk are
taken to be completely occulted, so $\tau \rightarrow \infty$.  The
other free parameter for the eclipse event is the trajectory of the
disk.  The trajectory is actually comprised of two factors: (1) the
impact parameter of the disk center relative to the star center in
the plane of the sky, and (2) the orientation of the star's spin
axis relative to that trajectory.  It is the latter that relates
to the fact that the eclipse leads to a significant excursion of
the polarization in the direction of negative Stokes-$u$.

\begin{figure*}
\plotone{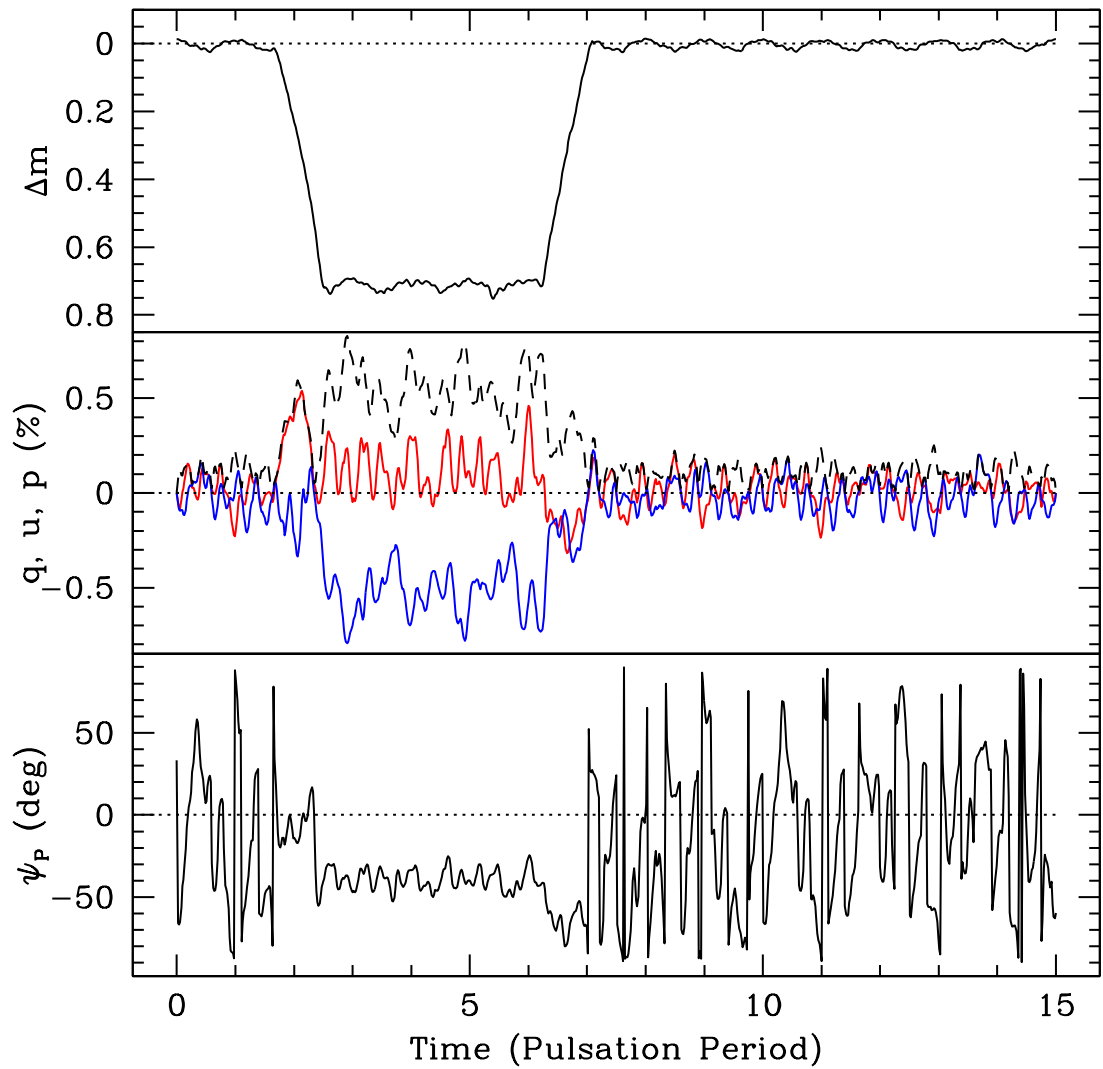}
\caption{Model for the primary star and an eclipse by a disk about
the secondary star using (2,2) and (3,1) modes of equal strength.
Parameters for temperature and temperature variation are adopted
from the solution (green dot) indicated in Fig.~\ref{fig7}.  The
figure for the model replicates the style for the data displayed
in Fig.~\ref{fig1}.
\label{fig8}}
\end{figure*}

Observations indicate that the disk runs across the lower half of
the star during the eclipse event.  This results in a drop of
brightness of about 0.7 magnitudes in $V$.  When running the disk
parallel to the equator of the star, an excursion of the polarization,
relative to outside of eclipse, does occur, but primarily in the
direction of $q$.  In order to obtain the observed excursion primarily
in the direction of $u$, the star's spin axis must be inclined
significantly.  While we take the F~star as seen edge-on by observer, and we introduce an obliquity angle on the plane of the sky, $\psi_\ast$ between the spin axis and the normal to the disk. A value of $\psi_\ast = 65^\circ$.

\begin{figure}
\plotone{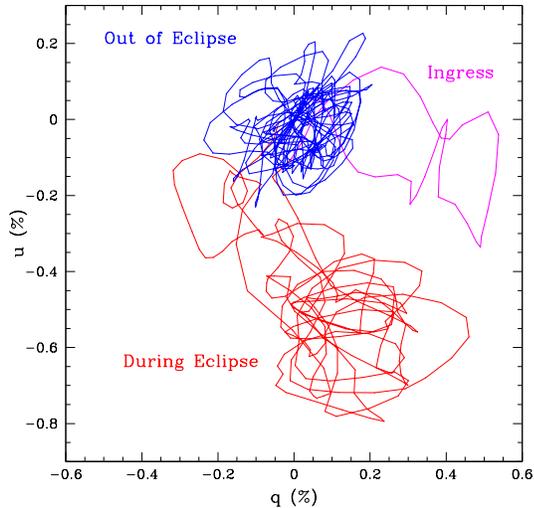}
\caption{The is the $q-u$ plane for the model shown in Fig.~\ref{fig8}.
It uses the same style as for the data displayed in Fig.~\ref{fig2}.
The only difference is that $\bar{q}=\bar{u}=0$ for the model.  The different colors are as labeled.  Specifically, ingress is identified in magenta since that portion of the eclipse was part of the model but does not have a counterpart in the observational dataset obtained by \cite{1989PhDT........11H}, which began after ingress.  
\label{fig9}}
\end{figure}

In the spirit of proof-of-concept, Figures~\ref{fig8} and \ref{fig9}
display model results that mimic the styles of Figures~\ref{fig1}
and \ref{fig2} for the observations reported by Henson (1989).  The
model results are for a disk that is $16R_1$ in diameter, with a
height that is $0.75R_1$ as seen edge-on.  The top of the disk, as
projected against the F~star, is $0.1R_1$ above the star
center; the bottom of the disk stretches to $0.65R_1$ below star
center.  The periods involved are listed in Table~\ref{tab2}. 
Mode switching (i.e., phase drift) occurs
at the two relevant angular beat frequencies for $l=2$ and $l=3$,
as previously discussed.

\begin{deluxetable}{lc}
\tablecaption{Model Parameters for $\epsilon$ Aur\label{tab2}}
\tablehead{
\colhead{Property} & \colhead{Value} 
}
\startdata
$T_0$ & 7,450~K \\
$\Delta T^4/T_0^4$ & 0.33 \\
$\gamma_0$ & 0.08 \\
$\psi_\ast$ & $65^\circ$ \\
$P_\ast$ & 1.0 years \\
$P_0$ & 0.30 years \\
$P_2$ & 0.17 years \\
$P_3$ & 0.094 years \\
\enddata
\end{deluxetable}

Notable points for Figure~\ref{fig8} include the amount of eclipse
in the light curve at
approximate correct depth, along with variations in brightness at
the level of a tenth of a magnitude (uppermost panel).  From the
middle panel, the $q$ and $u$ polarizations show complex behavior
outside of eclipse.  During eclipse, the $u$ polarization is notably
shifted toward negative values.  The $q$ polarization tends to be net
positive, but can dip to small negative values as observed, and shows an excursion
to the negative during egress.  The polarization position angle
(lowest panel) displays rapid position angle rotations outside eclipse,
but hovers around $-50^\circ$ throughout the eclipse.  These are
all properties very like the observations.

Figure~\ref{fig9} shows the model results in a $q-u$ plot.  The
blue portion is for variations out of eclipse; the red is during
the eclipse, with magenta corresponding to the ingress portion of
the eclipse.  The blue portion shows smoothly varying changes that
are mostly randomly oriented.  During eclipse, the polarization
becomes larger, with a net excursion from the blue toward negative
values of $u$.  At the same time, there continue to be smoothly
varying changes that are essentially centered on an offset position
from the blue portion.  This too is fairly similar to the observations.
We note that the magenta portion is not represented in the data
acquired by  \cite{1989PhDT........11H}, as those observations began
during the 1980's eclipse, but after the ingress had passed.  However,
the observations obtained during the 2009-2011 eclipse by
\cite{2012AIPC.1429..140H} do indicate an excursion predominantly
in $q$ during ingress.  This gives additional support to the model
reproducing the overall characteristics of the eclipse polarization.

One important note should be made about the modeling.  The excursion
to negative values of $u$ during eclipse requires that the spin
axis of the star (from which the function $G$ for the temperature
variations are defined) is inclined to the axis for the orbital
plane of the binary.  However, our modeling approach is meant to
identify a size, amplitude, and pattern of brightness variations
across the F~star that can match the observed polarimetric data.
The ``spin axis'' and mode switching serves merely to allow the
brightness pattern to evolve.  It is perhaps better to focus on the
size scale of the surface variations and the timescale for how the
pattern evolves.  These are properties that could be tested against
future interferometric observations\footnote{A ``white paper''
by Roettenbacher et al.\ (2019, arXiv:1903.04660) addresses prospects
for resolving stellar surface features at high angular resolution.}.


\section{Summary} \label{sec:summary}

The match between the data and the model is by no means exact.  It
was not our intention to produce an actual fit to the data.  What
has been accomplished is that the characteristic observed variations
(in shape and in amplitude) are broadly reproducible with our
phenomenological model that is motivated by plausible expectations
for the stellar atmosphere.  Our model does have some degeneracies,
such as the choice of limb polarization, and the amplitude for the
time-dependence of the average surface temperature.  More physically
motivated, and therefore more numerically intensive, stellar
atmosphere calculations are needed to obtain superior models in
order to understand the variable nature of the F~star.  Those models
will have consequences for better interpreting the eclipse event,
with ramifications for understanding the disk and perhaps the history
of this unusual system.

We point to one particular assumption that might be relaxed.  Our
model for brightness variations makes use of a temperature amplitude.
The value needed to match the observations was $\Delta T^4/T_0^4 =
0.33$.  At $T_0 = 7,450$~K, the implied range of temperatures across
the star is rather large at 6,740--8,000~K.  However, one assumption
of the model is a fixed radius.  Detailed atmosphere models reveal
that late-type supergiants can deviate from spherical owing to
convection.  If such effects exist in the F~star of $\epsilon$~Aur,
then the distorted shape of the star alone would itself produce a
net polarization.  Combined with stellar rotation, the F~star would
display variable polarization even without temperature variations.
Allowing for a variable shape of the star could reduce the amplitude
of brightness variations (and in our model, temperature variations)
across the atmosphere.

What does seem clear from our simplified approach is that (a) complex
surface brightness variations are likely needed to explain the
observed variable polarization data, (b) some modulation of the
stellar temperature is likely needed to explain the observed color
variations, and (c) and the distribution of brightness variations
must evolve over time.  These are features could be tested against
new interferometric observations for better understanding the F~star
component of the $\epsilon$~Aurigae system.

\acknowledgments

The authors are grateful to an anonymous
referee whose comments have improved this manuscript.
Ignace acknowledges funding support for this research from a grant
by the National Science Foundation, AST-2009412.  Henson acknowledges
funding support for this research from a grant by the National
Science Foundation, AST-1747658.  We acknowledge with thanks the
variable star observations from the AAVSO International Database
contributed by observers worldwide and used in this research.




\bibliography{epsAur}

\end{document}